\documentclass[prb,twocolumn,showpacs,floats,eqsecnum,amsmath,amssymb]{revtex4}
\usepackage[dvips]{graphicx}
\begin{document}


\title {Rotating Bose-Einstein condensate in an optical lattice: formulation of vortex configuration for the ground state}

\author {Y. Azizi and A. Valizadeh}

\affiliation{ Institute for Advanced Studies in Basic Sciences,
Zanjan 45195-1159, Iran}
\begin{abstract}
We consider a rotating Bose-Einstein condensate in an optical lattice in the regime in which the system Hamiltonian
can be mapped onto a Josephson junction array. In an approximate scheme where the couplings are assumed uniform,
the ground state energy is formulated in terms of vortex configuration. Application of method for ladder case presented
and the results are compared with Monte-Carlo method.
\end{abstract}
 \maketitle
\section{Introduction}

After the first experimental realization of the Bose-Einstein
Condensates (BECs)\cite{R33}, this field and its related topics
attracted more attentions\cite{R40,R41}. BEC in an optical
lattice \cite{R5,R29} has relation with many problems in the
condensed matter physics, e.g. Bloch
Oscillations\cite{R6,R7,R8,R9}, Wannier-Stark Ladders\cite{R10},
Josephson junction arrays\cite{R1,R3,R11,R25,R28,R48,R49} and
superfluid to the Mott-insulator
transition\cite{R12,R14,R22,R30,R51}.

Behavior of a rotating BEC in an optical lattice, is similar to
that of a superconductor in the magnetic field\cite{R15,R16,R17}.
Here with a trap potential without the lattice structure, the
Abrikosov vortex lattice can be
observed\cite{R1,R3,R25,R28,R18,R19,R20,R21,R26,R27}. When the
lattice structure adds to the trap potential, the vortex lattice
changes \cite{R50} and with some criterions, this study can be
mapped on the problem of the Josephson junction arrays (JJAs);
there will be the same vortex structure for both
systems\cite{R3,R25}. Most of the studies in this field has been
done on the lattice potentials with the square symmetries and few
works on the BEC with other symmetries\cite{R2,R13}. Yet due to
the possibility of the realization of the optical lattices with
different symmetries \cite{R4}and with more than one spatial
frequency\cite{R34}, the study of the BEC in the quasiperiodic
potentials can be useful\cite{R2,R13}.

Here we focus on the study of the problem in the JJA regime.
We formulate the problem of the ground state for the JJA Hamiltonian in terms of
the vortex configuration and then we discuss about the exact numerical result for the
few number of the lattice points, and the Monte-Carlo method for a large lattice.
We exploit harmonic approximation for the cosine Hamiltonian of the system and
find energy for any given vortex configuration. Also we study the problem for the ladder case;
this case can be used as the reduced two dimensional case\cite{R32} and
can inspire the general behavior of the ground state structure in the 2D lattice case.
This method is faster than the Monte-Carlo method for the original Hamiltonian and directly results the vortex lattice,
while in other methods instead of the vortex, the circulation has been used for the determination of the vortex position\cite{R42}.

\section{Rotating BEC in an optical lattice}
In this section we show how in some approximation,
rotating BEC in an optical lattice can be modeled as Josephson
junction arrays\cite{R1,R3,R25}. Hamiltonian for a BEC in an optical
lattice with a rotation frequency $\Omega \hat{z}$ is,
\begin{equation}\label{bechami}
\hat{H}=\int dr \hat{\Psi}^\dagger[\frac{(-i\hbar \nabla - m\Omega
\hat{z}\times r)}{2m}+V_{ext}+\frac{g}{2}\hat{\Psi}^\dag-\mu
]\hat{\Psi}
\end{equation}
where $m$ is the atomic mass and $g=4\pi \hbar^2a/m$ the coupling
constant with the $s$-wave scattering length $a$. Conservation of
the total number of particles is ensured by chemical potential
$\mu$. The external potential $V_{ext}=V_h+V_O$ consists of two
parts: modified harmonic potential
$V_h=m(\omega_{\perp}^2-\Omega^2)r^2/2+m\omega_z^2z^2/2$ and the
potential of the optical lattice $V_O$, which may be chosen as
periodic\cite{R36}, quasiperiodic\cite{R13,R2}, or
random\cite{R35}. For example, the potential with square symmetry
can be written as $V_O=V_0[sin^2(kx)+sin^2(ky)]$, with the
periodicity $\pi/k$.


For a large $\omega_z$, we can suppose that system is frozen in axial direction. If energy due
to interaction and rotation is small compared to the energy
separation between the lowest and first excited band, the
particles are confined to the lowest Wannier orbitals\cite{R25}. Therefore We can write $\psi$ in the
Wannier basis as,

\begin{equation}\label{wannier}
\hat{\Psi}(r)=\sum_{i=1}^N\hat{a}_iw_i(r)exp(\frac{im}{\hbar}\int_{r_i}^r
A(r').dr', )
\end{equation}
where $A=\Omega\hat{z}\times r$ is the analog of the magnetic
vector potential, $w_j(r)=w(r-r_i)$ the normalized Wannier wave-function
localized in the $i$-th well and $\hat{n}_i=\hat{a}_i^{\dag}\hat{a}_i$ the number operator. Substituting \ref{wannier} in Hamiltonian
\ref{bechami} leads to the Bose-Hubbard model in the rotating frame

\begin{eqnarray}\label{bosehubbard}
\hat{H}&=&-t\sum_{<i,j>}(\hat{a}^{\dag}_i\hat{a}_je^{-A_{ij}}+h.c.)\nonumber\\
&+&\sum_i(\epsilon_i-\mu)\hat{n}_i+\frac{U}{2}\sum_i\hat{n}_i(\hat{n}_i-1),
\end{eqnarray}
where $\langle i,j\rangle$ means $i$ and $j$ are nearest neighbors, $t\approx
-\int dr w_{j}^*(-\hbar^2\nabla^2/2m+V_O)w_j $ is the hopping matrix
element, $\epsilon_i=\int dr w^*_iV_hw_i$ an energy offset of
each lattice site and $U=g\int dr|w_i|^4$ the on-site energy. The
rotation effect is described by $A_{ij}$ proportional to the
line integral of $A$ between the $i$-th and $j$-th sites:
$A_{ij}=(m/\hbar)\int_{r_i}^{r_{j}}A(r').dr'$.

\begin{figure}[ht!]
\centerline{\includegraphics[width=10cm]{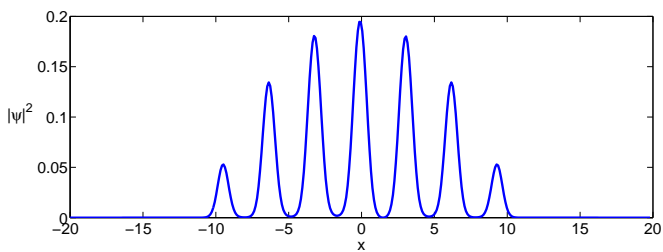}}
\vspace{0.1cm}
\caption{Solution of GP equation for a periodic one dimensional potential. It can be seen that $|\Psi|^2$ is enveloped by an inverted parabola shape.}
\vspace{-0.0cm} \label{oned}
\end{figure}

If the number of atoms per sites is large ($n_i\gg 1$), the
operator can be expressed in terms of its amplitude and phase,
and the amplitude by the $c$-number as $\hat{a_i}\approx \sqrt{n_i}e^{i\hat{\theta}_i}$.
Using $\hat{n}_i=n_i-i\partial/\partial\theta_i$ and $\hat{\theta}_i=\theta_{i}$,
Eq. \ref{bosehubbard} reduces to the quantum phase model
\begin{equation}
\hat{H}=-\sum_{<i,j>}J_{ij}cos(\theta_i-\theta_{j}+A_{ij})-\frac{U}{2}\sum_i\frac{\partial^2}{\partial\theta_i^2},
\end{equation}
where $J_{ij}=2t\sqrt{n_in_{j}}$. Also, the atoms are
assumed to be distributed such as they satisfy the condition
$\epsilon_i+Un_i=\mu$. The magnitude of $J_{ij}$ decreases from
the central sites outward because $n_i$ has the profile of
an inverted parabola as we can see from solution of 1D case in Fig. \ref{oned}.

Eq. 2.4 is just the Hamiltonian of a lattice of small Josephson junctions with inhomogeneous coupling constants $J_{ij}$. With $J_{ij}\gg U$, we can neglect the kinetic energy term; Moreover, we suppose that coupling constants are equal
$J_{ij}=J$. This regime can be achieved when the depth of optical
lattice $V_0$ is between 18 and 25$E_R$ ($E_R=\hbar^2k^2/2m$ is
recoil energy) and average particle number is $\bar{n}\approx
170$\cite{R31}. Then we have:
\begin{equation}\label{hamiltonian}
H=-\sum_{<i,j>}cos(\gamma_{ij}),
\end{equation}
with $\gamma_{ij}=\theta_{i}-\theta_{j}-A_{ij}$. Our purpose is
to find minimum of this Hamiltonian as a function of
condensate phases $\theta_{i}$ with two constraints as follows.
First constraint originates in the single valuedness of the wave
function which leads to the quantization of vorticity:
\begin{equation}
\sum \gamma_{ij}=2\pi(n_k-a_kf)
\end{equation}
where sum is over $k$th plaquette, $a_k$ is area of $k$th
plaquette, and $n_k$ is integer. In the case of a condensate with the
rigid-body rotation, {\it the frustration parameter} is defined as $f=2\Omega/\kappa$
with the quantum circulation $\kappa=h/m$. In two dimensional arrays, frustration represents
density of vortices per number of lattice plaquettes \cite{R44}.

The second constraint comes from conservation of the particles numbers.
The particle current from the nodes can be assumed as the derivation of the energy function
with respect to the corresponding phase\cite{R43}. This leads to
\begin{equation}\label{ccons}
\sum sin(\gamma_{ij})=0,
\end{equation}
where sum is over all $j$ which is connected to $i$th node.

Since Hamiltonian is periodic, $n_k$ can be $\lfloor
a_kf\rfloor$ or $\lfloor a_kf\rfloor+1$, and we can replace
them by $n_k$ which is zero or one, and replace $a_kf$ with
$a_kf-\lfloor a_kf\rfloor$. These two set of equations can
completely determine all $\gamma_{ij}$ for a given set of $n_k$.
Also, for a lattice, number of plaquettes plus number of nodes is equal to number of
connections minus one\footnote{This is Euler formula for polygon lattice excluding exterior face.};
therefore one of the equations (e.g. one of particle conservation equations) can be neglected.
Assuming the particle flow from the nodes are small, we can linearize the Eq.\ref{ccons}
i.e. $sin{\gamma_{ij}\approx \gamma_{ij}}$. Now, we have a linear set of equations which must
be solved for a given set of integers $n_k$, i.e. for a vortex configuration. Energy of this vortex
configuration is denoted by $E_{\{n\}}$. Then instead of original minimization scheme, we can use the set of these
energies for minimization of Hamiltonian. It means that minimum
of Hamiltonian correspond to a vortex configuration which minimizes these set of energies,
with number of vortex configuration equal to $2^{N}$ where $N$ is number of plaquettes in the lattice.

\section{Ladder case}
In this section, we apply the formulation of previous section on a
ladder geometry and find an exact analytical formula for energy of vortex configuration.
It is known that a ladder geometry can give an insight about ground state vortex structure
in 2D lattices\cite{R32}. As we will see the symmetry of the problem in this case,
reduces the number of equations such that there will be $N$ linear algebraic equations to solve,
where $N$ is number of plaquettes. One of difference between problems in Josephson junction arrays
and BEC lattices is boundary condition which is imposed on lattice. In the Josephson junction arrays,
the behavior of the arrays with large number of plaquettes is interesting\cite{R44},
therefore the boundary condition of lattice is usually assumed as periodic,
i.e. the first plaquette is connected to the last plaquette.
For BECs, the number of lattice sites is usually finite and free or fixed boundary
condition are more suitable.

Conservation of current \ref{ccons} imposes that current in upper and lower junctions of a plaquette and therefore their corresponding $\gamma_{ij}$ will be equal. We denote gauge invariant phase difference for upper and lower junctions of $k$th plaquette by $\gamma_k$ and those of the vertical junctions between $k$ and $k+1$th plaquette by $\gamma_{k,k+1}$. Then the node current equations read
\begin{equation}
sin(\gamma_{k,k+1})=sin(\gamma_k)-sin(\gamma_{k+1}),
\end{equation}
and the linearized approximation gives

\begin{equation}
\gamma_{k,k+1}=\gamma_k-\gamma_{k+1}.
\end{equation}

Imposing this equation on the flux quantization 2.6, we have

\begin{equation}
4\gamma_k-\gamma_{k+1}-\gamma_{k-1}=2\pi(n_k+\lfloor a_kf\rfloor-a_kf),
\end{equation}
for the $k$th plaquette with fixed boundary condition: $\gamma_{N+1}=\gamma_0=0$, where $N$ is number of plaquettes in ladder. We have left now with $N$ equations and $2N$ variables. Denoting coefficient matrix by $C$, we can write

\begin{equation}\label{gammaladder}
\gamma_k=\sum_{k=1}^NC^{-1}_{kl}b_l=2\pi(N_k-A_kf),
\end{equation}
where $b_l=2\pi(n_l-a_lf)$, and $C^{-1}$ is inverse of coefficient
matrix and its elements are,
\begin{equation}
C^{-1}_{kl}=\frac{\xi^{N-|k-l|+1}+\xi^{|k-l|-N-1}-\xi^{N+1-k-l}-\xi^{k+l-N-1}}{(\xi-\xi^{-1})(\xi^{N+1}-\xi^{-N-1})},
\end{equation}
where $\xi=2+\sqrt{3}$, and $A_k$ and $N_k$ are defined with appropriate summation.

By using Eq. \ref{gammaladder} we can find the energy (normalized by the number of junctions) for a vortex
configuration $\{n\}$ as:
\begin{eqnarray}\label{energynladder}
\nonumber
E_{\{n\}}&=&\frac{1}{3N+1}\{\sum_{k=1}^N\{2cos(\gamma_k)+cos(\gamma_k-\gamma_{k+1})\}\\
&+&cos(\gamma_N)\}.
\end{eqnarray}

\begin{figure}[ht!]
\centerline{\includegraphics[width=9cm]{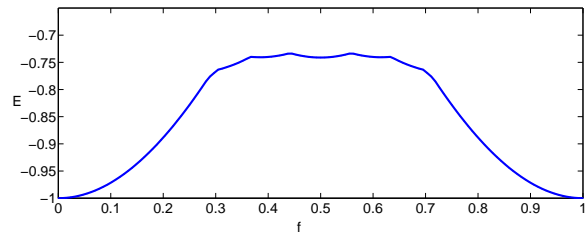}}
\vspace{0.1cm}
\caption{Ground state energy of a square lattice with 8 plaquettes. For each value of $f$, energy of all vortex configurations is calculated and the smallest value has been chosen.}
\vspace{-0.0cm} \label{sq8en}
\end{figure}

Related to the problem of minimizing the Hamiltonian, each vortex configuration can be
vortex configuration with minimum energy for a definite value, or an interval of $f$. Then for each value of frustration we need calculate the energy for all the vortex configurations. As an example, we begin with a square ladder ($a_i=1$) with $8$ plaquette in which we just need to study $0\le f<1$. In Fig. \ref{sq8en} we have plotted ground state energy by the above procedure: $f$ is incremented from $0$ to $1$ by $1/128$ and for each value of $f$, energy of all the configurations is calculated and the least value has been chosen. We have specified each vortex configuration by a number $m$ which in the binary representation with $N$ digit ($N$ is number of plaquettes) gives the configuration of the vortices with every one, meaning there is vortex on that plaquette. For example $m=3$ means a vortex configuration $00000011$, i.e. there are two vortices on the $7$th and $8$th plaquettes.

\begin{figure}[ht!]
\centerline{\includegraphics[width=9cm]{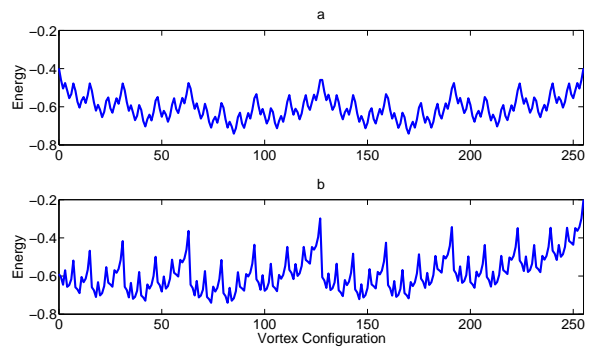}}
\vspace{0.1cm}
\caption{Energy of vortex configurations for square ladder with 8 plaquettes. Binary representation is used to demonstrate vortex configurations as is explained in the text. $f=1/2$ in (a) and $f=2/5$ in (b).}
\vspace{-0.0cm} \label{sq8vet}
\end{figure}

\begin{figure}[ht!]
\centerline{\includegraphics[width=10cm]{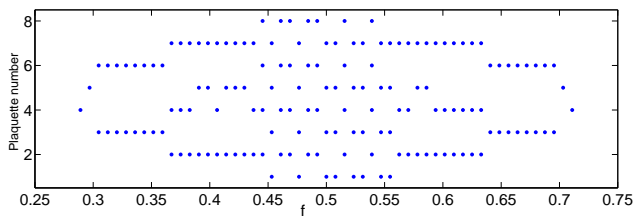}}
\vspace{0.1cm}
 \caption{Position of vortex for a square ladder with 8 plaquettes. Vertical axis shows the position of plaquette in the array.}
\vspace{-0.0cm} \label{sq8vp}
\end{figure}

Fixing $f$ we can demonstrate the dependence of the energy to the configuration of the vortices. In Fig. \ref{sq8vet} we have plot energy vs. vortex configuration for $f=1/2$ and $2/5$ where the vortex configuration is labeled as we noted above. For $f=1/2$ we see a mirror symmetry, it means that vortex configurations $m$ and $2^N-m$ have the same energy. More general rule for square ladder with arbitrary $f$ can be deduced easily from above equations: vortex configuration $m$ for $f$ and $2^N-m$ for $1-f$ has same energy. Therefore, if vortex configuration $m$ has the minimum energy for $f$, then vortex configuration $2^N-m$ has the minimum energy for $1-f$. This means that for $f=1/2$, the minimum energy is twofold degenerate. For example for the ladder considered above, both $01010101$ and $10101010$ configurations give the minimum energy.
In Fig. \ref{sq8vp} we have shown position of vortices for different $f$. This figure shows that behavior of vortex is similar to the behavior of a charge $q$ on a lattice with opposite sign onsite charges\cite{R44,R45}. When we have one vortex on lattice, the vortex prefers to sit on middle of ladder, where in this case it can be two plaquettes 4 or 5. When we have two vortices on ladder, they prefer to divide ladder into three equal parts, which means that they must be on plaquettes 3 and 6. This description can be applied to vortex configurations with more vortices.

\begin{figure}[ht!]
\centerline{\includegraphics[width=10cm]{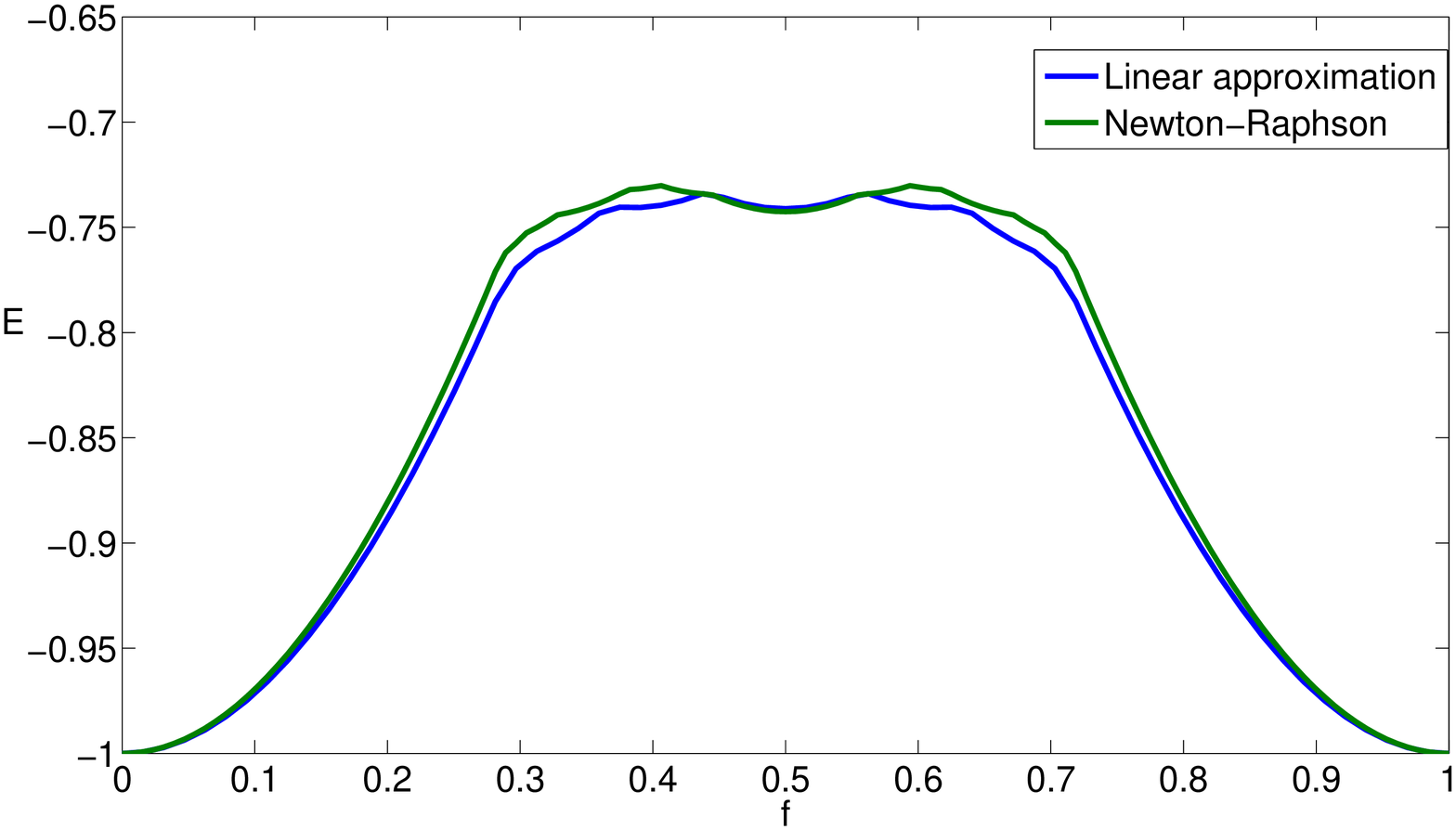}}
\vspace{0.1cm}
 \caption{Comparison between linear approximation (blue) and the Newton-Raphson method (green).}
\vspace{-0.0cm} \label{Linr}
\end{figure}

For more accurate results, we can relax linear approximation for the current equation and use Newton-Raphson method, with the initial estimation from linear approximation. For above case, we compare linear approximation and 10 iterates of Newton-Raphson method in Fig. \ref{Linr}. It worth to mention while energy differences are small but, in some small intervals of $f$, different vortex configurations give the minimum energy.

\section{MC formulation for vortex configuration}
When $N$ number of plaquettes grows, number of possible vortex
configurations grows as $2^N$, and expect for small number
of plaquettes we can not determine the exact minimum vortex configuration which needs checking all of vortex configuration. Therefore we use a Monte-Carlo method: we start from a high temperature $T$ and a random vortex lattice $n$ and calculate its energy $E_{n}$. Then we decrease temperature gradually and for each temperature we do following process a few times: we change the configuration randomly to $n'$ with energy $E_{n'}$, we accept the new configuration if $exp(-\Delta E/T)<r$ where $r$ is a random number between $0$ and $1$, and $\Delta E=E_{n}-E_{n'}$. When temperature receives $0.0001$, we take the final configuration as ground state.

\begin{figure}[ht!]
\centerline{\includegraphics[width=10cm]{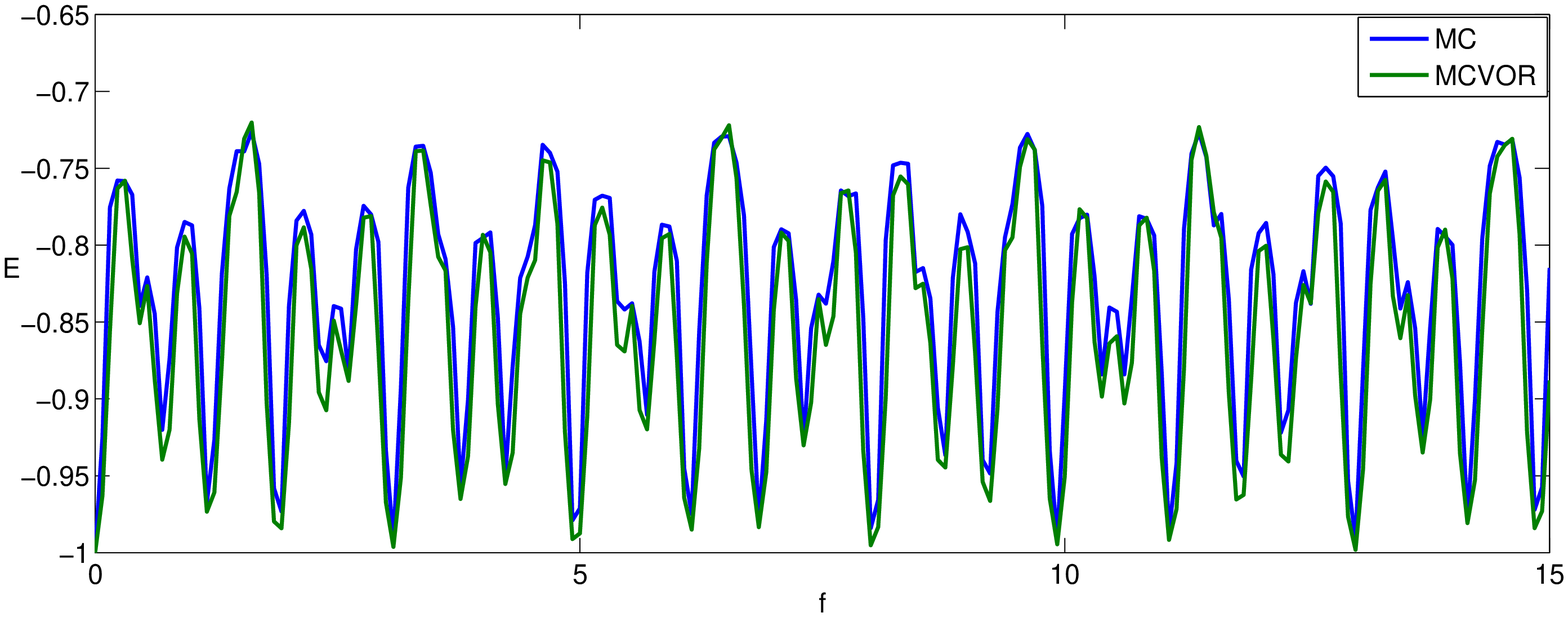}}
\vspace{0.1cm}
 \caption{Comparison between energy obtained by the direct Monte-Carlo method (MC) and by the Monte-Carlo method with vortex configuration formulation (MCVOR).}
\vspace{-0.5cm} \label{comparison}
\end{figure}

For a periodic lattice finding the ground state from the MC
method is troublesome for nonzero frustration\cite{R47}. To check
the validity of our method, instead, we apply the method on a
quasiperiodic lattice e.g. a Fibonacci ladder. The Fibonacci
ladders contains two types of plaquettes with lengths 1 and
$\tau=(1+\sqrt(5))/2$. If we denote plaquette with length 1 and
$\sigma$, by $S$ and $L$ respectively, then these two types of
plaquettes arranged in bases of the following rule: for $(n+1)$th
step of construction, we add $(n-1)$th step to end of $n$th step.
With first step $S$ and second step $L$, we have:
$LS,LSL,LSLLS,LSLLSLSL,...$. The results are shown in Fig.
\ref{comparison} along with the result of using the Monte-Carlo
on the original Hamiltonian, with a good agreement (we note that
for a aperiodic lattice it is not sufficient to study $0\leq f
<1$). For a square lattice, increasing $f$ from $0$ to $1/2$, both
the number of the vortices and the magnitude of vorticity grow
\cite{R44} and the linear approximation may seem unreasonable.
Fig. \ref{comparison} shows that despite this fact, the
approximate results are in a good agreement with the results of
the direct method.

%

\section{Conclusion}
The configuration of the vortex lattice of rotating BECs in an Optical
lattice is investigated. For deep Optical Lattices when the number of particles in each site is large, the problem of rotating BEC in an
optical lattice, can be mapped onto the model of arrays of Josephson junctions in presence of an external magnetic field. In this approximation we have formulated a vortex configuration for the ground state. Our result for ladder case is
presented and is in good agreement with Monte-Carlo result with original JJA Hamiltonian. This method can be extended to 2D case, and also cases with non-uniform coupling which looks more relevant for study of rotating BECs. In these cases, with linear approximation, we deal with a set of linear algebraic equations for each vortex configuration which can be solved easily by use of proposed method.

Effect of non-uniform coupling can affect the results especially in the situation of high density of vortices as in the case $f=1/2$ for square lattice. Also as a better approximation we can consider finite $U$, then problem is similar to the arrays of small Josephson junctions\cite{R37,R38,R39}. We can also suppose that coupling is time-dependent which can occur in case of a vibrating optical lattice\cite{R46}. Then problem can be treated as a set of coupled pendulum equations with time-dependent lengths\cite{R1}.


\end{document}